\documentclass[showpacs,preprintnumbers,amsmath,amssymb,12pt,floatfix,epsfig]{revtex4}
\usepackage{graphicx}
\begin{document}
\draft
\title{Ambiguities of neutrino(antineutrino) scattering on the
nucleon due to the uncertainties of relevant strangeness form
factors}
\author{Myung-Ki Cheoun$^{1)}$
\footnote{cheoun@ssu.ac.kr}, K. S. Kim$^{2)}$ }
\address{ 1)Department of Physics,
Soongsil University, Seoul, 156-743, Korea \\ 2)School of Liberal
Arts and Science, Korea Aerospace University, Koyang 200-1, Korea
}

\begin{abstract}
Strange quark contributions to neutrino(antineutrino) scattering
are investigated on the nucleon level in the quasi-elastic region.
The incident energy range between 500 MeV and 1.0 GeV is used for
the scattering. All of the physical observable by the scattering
are investigated within available experimental and theoretical
results for the strangeness form factors of the nucleon. In
specific, a newly combined data of parity violating electron
scattering and neutrino scattering is exploited. Feasible
quantities to be explored for the strangeness contents are
discussed for the application to neutrino-nucleus scattering.
\end{abstract}
\narrowtext
\maketitle

Knowledge of neutrino-nucleus interactions plays vital roles in
nuclear detectors for the neutrino($\nu$) physics, such as $\nu$
oscillations and $\nu$ masses. Therefore neutrino-nucleus $ (\nu -
A)$ scattering has become to be widely interested in different
fields of physics such as astrophysics, cosmology, particle, and
nuclear physics. Not only the $\nu$ physics, but also the hadron
physics is closely related to the $ \nu - A$ scattering. In
particular, the scattering of neutrino and antineutrino$({\bar
\nu})$ on nuclei enables us to obtain some invaluable clues on the
strangeness contents of the nucleon. Along this line, Brookhaven
National Laboratory (BNL) \cite{brook} reported that a value of a
strange axial vector form factor of the nucleon, $G_A^s(Q^2 = 0)$,
does not have zero through the experimental data of $\nu ({\bar
\nu})$ scattering on the proton. But the extraction of the exact
value $G_A^s (0)$ depends on other variables, such as the axial
mass, {\it i.e.}, the axial vector dipole mass, and the parameters
on strange vector form factors \cite{garvey,Albe02}.

Unfortunately, the BNL data is the only one for $\nu ({\bar \nu})$
scattering on the nucleon available until now, so that one uses
complementary data from the parity violating (PV) polarized
electron scattering by HAPPEX\cite{Aniol04} and $G^0$
data\cite{Arms05} in searches for the strangeness on the nucleon
\cite{Budd03}. Of course, there are on going experiments, FINeSSE
\cite{Fin05}, or proposed experiments in J-PARC \cite{Fili01}.

On the other hand, many calculations
\cite{umino,alberico,jacowicz,giusti1,udias,madrid,kim07} about
$\nu - A$ scattering are carried out to disentangle effects of the
strangeness from $\nu$ interactions with exotic nuclei matter,
which is one of the key ingredients of understanding the supernova
explosion. But there remained still some uncertainties from the
undetermined strangeness form factors of the nucleon to be pinned
down for further nuclear application.

Recently, however, Pete \cite{Step04} suggested a method to
combine $\nu({\bar \nu})$ - nucleon(N) data and PV electron
scattering data, and displayed a data set composed of 11 data for
the strangeness form factors \cite{Step06}. Therefore, by
exploiting the new data set, it is necessary and meaningful to
investigate again the ambiguities regarding the interpretation of
$\nu - N$ scattering data, which stems from the uncertainties of
strangeness form factors. This could constrain the uncertainties
relevant to the strangeness on the nucleon and give more
reasonable analysis for further study of the $\nu -A$ scattering.

In this paper, we consider the neutral current (NC) and charged
current (CC) scattering on the nucleon within the relevant data in
the quasi-elastic (QE) region, where inelastic processes like pion
production and $\Delta$ resonance are excluded. Beyond the QE
region, one has to include such inelastic processes. For example,
ref. \cite{leitner} showed that the contribution of $\Delta$
excitation is comparable to that of the QE scattering at the
neutrino energies above 1 GeV.

We start from a weak current on the nucleon level. The weak
current, $W^{\mu}$, takes a $V^{\nu} - A^{\nu}$ current form by
the standard electro-weak theory, which has isoscalar and
isovector parts for NC interactions
\begin{eqnarray}
W^{\mu} & = & V_{3}^{\mu} - A_{3}^{\mu} - 2 {sin}^2
{\theta}_{W} J_{em}^{\mu} - { 1 \over 2} ( V_s^{\mu} - A_s^{\mu}) \\
\nonumber & = & ( 1 - 2 {sin}^2 {\theta}_{W} ) V_{3}^{\mu} -
A_{3}^{\mu} - 2 {sin}^2 {\theta}_{W} V_{0}^{\mu} - { 1 \over 2} (
V_s^{\mu} - A_s^{\mu}) ~,
\end{eqnarray}
with Weinberg angle $\theta_W$, where we used $J_{em}^{\mu} =
V_3^{\mu} + V_0^{\mu}$. Strangeness contributions, which are
isoscalar parts, are considered in $ - { 1 \over 2} V_{s}^{\mu}+ {
1 \over 2} A_{s}^{\mu}$. For CC interactions, only $V_{3}^{\mu} -
A_{3}^{\mu}$ term is involved, while $J_{\mu}^{em} = V_{3}^{\mu} +
V_{0}^{\mu}$ is concerned with the meson electro-production.
Therefore the CC scattering of $\nu ({\bar \nu})$ is independent
of strangeness contents. For the elastic scattering of polarized
electron on nucleon, $J^{\mu} = - 2 {sin}^2 {\theta}_{W}
J_{em}^{\mu} - { 1 \over 2} V_s^{\mu} $ is exploited.

For a free nucleon, the current operator comprises the vector and
the axial vector form factors, $F_i^V (Q^2)$ and $G_A (Q^2)$,
\begin{equation}
{W}^{\mu}=F_{1}^V (Q^2){\gamma}^{\mu}+ F_{2}^V (Q^2){\frac {i}
{2M_N}}{\sigma}^{\mu\nu}q_{\nu} + G_A(Q^2) \gamma^{\mu} \gamma^5 +
{ G_P(Q^2) \over {2M}} q^{\mu} \gamma^5.
\end{equation}
By the conservation of the vector current (CVC) hypothesis with
the inclusion of an isoscalar strange quark contribution, $F_i^s$,
the vector form factors for protons and neutrons, $F_{i}^{V,~p(n)}
(Q^2)$, are expressed as \cite{giusti1}
\begin{eqnarray}
F_i^{V,p(n)} ( Q^2) &=&({\frac 1 2} - 2 \sin^2 \theta_W )
F_i^{p(n)} ( Q^2) - {\frac 1 2} F_i^{n(p)}( Q^2) -{\frac 1 2}
F_i^s ( Q^2)~~ for~ NC\\ \nonumber & =& ( F_i^{p} ( Q^2) -
 F_i^{n}( Q^2))~~for~ CC~.
\end{eqnarray}
Strange vector form factors are usually given as a dipole
form\cite{garvey}, independently of the nucleon isospin,
\begin{equation}
F_1^s(Q^2) = {\frac {F_1^s Q^2} {(1+\tau)(1+Q^2/M_V^2)^2}}~,
\;\;\;\;\; F_2^s(Q^2) = {\frac {F_2^s(0)}
{(1+\tau)(1+Q^2/M_V^2)^2}}~,
\end{equation}
where $\tau=Q^2/(4M_N^2)$, $M_V=0.843$ GeV is the cut off mass
parameter usually adopted for nucleon electro-magnetic form
factors. If we assume the same $Q^2$ dependence as the non-strange
Sachs form factors
\begin{equation}
G_E (Q^2) = F_1 (Q^2) - { {Q^2} \over { 4 M^2}} F_2 (Q^2)~,G_M
(Q^2) = F_1 (Q^2) + F_2 (Q^2)~,
\end{equation}
one obtains electro and magnetic strangeness form factors
\begin{equation}
G_M^s (Q^2) = {  {Q^2 F_1^s + \mu_s  } \over { ( 1 + \tau) {( 1 +
Q^2/M_V^2)}^2 }}, ~G_E^s (Q^2) = {  {Q^2 F_1^s - \mu_s \tau  }
\over { ( 1 + \tau) {( 1 + Q^2/M_V^2)}^2 }}~,
\end{equation}
where $\mu_s = F_2^s(0)=G_M^s (0) $ is a strange magnetic moment.
If we define the root means square (rms) value of strangeness,
$<r_s^2> = - 6~ {  {d G_E^s (Q^2) } / {d Q^2 }} \vert_{Q^2 = 0}$,
it can be approximated as $<r_s^2> \sim - 6 F_1^s$ at small $Q^2$.
Therefore, $F_1^s$ can be deduced from $G_{E}^s (Q^2)$
\cite{Step06}. There exists a bit different form from the above
one \cite{giusti1}.

The axial form factor is given by \cite{musolf}
\begin{eqnarray}
G_A (Q^2) &=&{\frac 1 2} (\mp g_A + g_A^s)/(1+Q^2/M_A^2)^2~~ for~
NC~\\ \nonumber G_A^{CC} (Q^2) & =& - g_A / {( 1 + Q^2 /
M_A^2)}^2~~ for~ CC~, \label{gs}
\end{eqnarray}
where $g_A$ and $M_A$ are the axial coupling constant, the axial
cut off mass, respectively. $-(+)$ coming from the isospin
dependence denotes the knocked-out proton (neutron), respectively.
The $g_A^s $ represents the strange quark contents in the nucleon
. Usually it is interpreted as the integral over the polarized
parton distribution function by the strangeness quark
\cite{Bern02}. Since we take $+$ sign for $G_A (Q^2) $ in Eq.(2),
the axial form factor in Eq.(7) is just negative to the form
factor elsewhere, for example, in ref.\cite{giusti1}. The induced
pseudoscalar form factor is usually parameterized by the
Goldberger-Treimann relation
\begin{equation}
G_P(Q^2) = {\frac {2M_N} {Q^2+m^2_{\pi}}} G_A(Q^2),
\end{equation}
where $m_{\pi}$ is the pion mass. But the contribution of the
pseudoscalar form factor vanishes for the NC reaction because of
the negligible final lepton mass participating in this reaction.

In principle, relevant strangeness form factors, $G_{E,M}^s (
Q^2)$ and $G_A^s (Q^2) = g_A^s / 2 {( 1 + Q^2/M_A^2)}^2$, can be
deduced from experimental data, for instance, $G^0$ and HAPPEX
data by PV electron scattering and/or BNL data by $\nu - N$
scattering. But the available experimental data are not enough to
fix the strangeness form factors, although theoretical
calculations, such as chiral soliton nucleon model or quark model
\cite{Silvia,Riska}, show results consistent, but scattered with
each other, with the experimental data. Therefore, values of
$F_1^s$, $\mu_s$, and $g_A^s$ in the form factors have some
ambiguities due to the uncertainties persisting in the strangeness
form factors \cite{Budd03,Step06}.

But the newly combined data set \cite{Step06} for those
strangeness form factors (see figure 1 at ref.\cite{Step06}) makes
it possible to constrain the relevant parameters to some extent by
adjusting the parameters to the data. In table 1, relevant
coupling constants and parameters adopted in previous calculations
are summarized with our values adjusted to the experimental data
\cite{Step06}.

Model IV in ref. \cite{garvey} deduced from the reanalysis of BNL
data is the most common value in $\nu - N (A) $ scattering
calculations, and difference between model III and IV is the axial
mass used. Ref. \cite{udias} did not take the strangeness into
account, but focus on the $\Delta$ region. We fixed $sin^2
\theta_W, g_A, M_A$ as the most common values. But other
parameters are adjusted to satisfy newly combined data
\cite{Step06} and theoretical calculations, by presuming a dipole
form. It is remarkable that $\mu^s$ value is constrained to be
positive according to the newly combined data. This contradicts
the previous values used in previous $\nu -A$ scattering
calculations.

Our lower and upper limits of the parameters constrained to
experimental data lead to some inescapable ambiguities of the
physical observales. They are discussed in detail in the
following. Since the first term in Eq.(3) rarely contributes due
to the given Weinberg angle, the elastic cross section on the
proton, $\sigma ( \nu p \rightarrow \nu p)$, is sensitive mainly
on the $F_i^s$ and $g_A$ values. But measuring of the cross
section itself is not easy experimental task, so that one usually
resorts to the cross section ratio between the proton and the
neutron, $R_{p/n} = { {\sigma ( \nu p \rightarrow \nu p ) } / {
\sigma (\nu n \rightarrow \nu n )}}$. Measuring of this ratio has
also some difficulties in the neutron detection.

Therefore, the ratio, $R_{NC/CC} = { {\sigma ( \nu n \rightarrow
\nu n ) } / { \sigma (\nu n \rightarrow \mu^- p )}}$ and ${\bar
R}_{NC/CC} = { {\sigma ( {\bar \nu} p \rightarrow {\bar \nu} p ) }
/ { \sigma ({\bar \nu} p \rightarrow \mu^+ n )}}$, are suggested
as plausible signals for the nucleon strangeness, for example, in
FINeSSE experiments \cite{Fin05}, because the charged current (CC)
cross section is insensitive to the strangeness \cite{Vent04}. In
this paper, we investigate other possible observables, mainly
asymmetries between $\nu$ and ${\bar \nu}$ scattering.

If we assume that the standard Sachs form factors also hold for
vector form factors
\begin{equation}
G_E^V (Q^2) = F_1^V (Q^2) - { {Q^2} \over { 4 M^2}} F_2^V
(Q^2)~,G_M^V (Q^2) = F_1^V (Q^2) + F_2^V (Q^2)~,
\end{equation}
we can define Sachs vector form factors, similarly to Eq.(3),
\begin{eqnarray}
G_{M,E}^{V,~ p(n)} ( Q^2) &=&({\frac 1 2} - 2 \sin^2 \theta_W )
G_{M,E}^{p(n)} ( Q^2) - {\frac 1 2} G_{M,E}^{n(p)}( Q^2) -{\frac 1
2} G_{M,E}^s ( Q^2)~~ for~ NC~\\ \nonumber & =& ( G_{M,E}^{p} (
Q^2) -
 G_{M,E}^{n}( Q^2))~~for~ CC~.
\end{eqnarray}

Then the NC $\nu {(\bar \nu})- N$ cross section is expressed in
terms of the Sachs vector and axial form factors \cite{Albe02}
\begin{eqnarray}
{( {  {d \sigma  } \over {d Q^2  }} )}_{\nu ({\bar \nu}   )
}^{NC}&=& { {G_F^2 } \over {2 \pi }} [ { 1 \over 2} y^2 { (G_M^V
)}^2 + ( 1 - y - { { M }\over {2 E_{\nu} }} y ) { {{ (G_E^V
)}^2 + { E_{\nu} \over 2 M} y { (G_M^V )}^2 } \over {1 + {  E_{\nu} \over 2 M} y }} \\
\nonumber & & + ( { 1 \over 2} y^2 + 1 - y + { { M } \over {2
E_{\nu} }} y ){ (G_A )}^2 \mp 2 y ( 1 - { 1 \over 2 } y ) G_M^V
G_A ]~.
\end{eqnarray}
Here $E_{\nu}$ is the energy of incident $\nu ({\bar \nu})$ in the
laboratory frame, and $y = { { p \cdot q } / { p \cdot k }} = { {
Q^2 } / {2 p \cdot k }}$ with $k,p$ and $q$, initial 4 momenta of
$\nu ({\bar \nu})$ and target nucleon, and 4 momentum transfer to
the nucleon, respectively. $\mp$ corresponds to the cases of the
$\nu$ and ${\bar \nu}$. Therefore the difference and the sum of
the cross sections are simply summarized as
\begin{equation}
{( {  {d \sigma  } \over {d Q^2  }} )}_{\nu     }^{NC} - {( { {d
\sigma  } \over {d Q^2  }} )}_{ {\bar \nu}    }^{NC} = - { {G_F^2
} \over { 2 \pi }}~ 4 y ( 1 - { 1 \over 2} y ) G_M^V G_A~,
\end{equation}
\begin{eqnarray}
{( {  {d \sigma  } \over {d Q^2  }} )}_{\nu     }^{NC} + {( { {d
\sigma  } \over {d Q^2  }} )}_{ {\bar \nu}    }^{NC} &=& { {G_F^2
} \over { \pi }} [ { 1 \over 2} y^2 { (G_M^V )}^2 + ( 1 - y - { {
M }\over {2 E_{\nu} }} y ) { {{ (G_E^V
)}^2 + { E_{\nu} \over 2 M} y { (G_M^V )}^2 } \over {1 + {  E_{\nu} \over 2 M} y }} \\
\nonumber & & + ( { 1 \over 2} y^2 + 1 - y + { { M } \over {2
E_{\nu} }} y ){ (G_A )}^2 ]~.
\end{eqnarray}
As shown in Eq.(12), the difference between $\nu$ and ${\bar \nu}$
scattering depends only on the product of $G_M^V$ and $G_A$.
Consequently, the asymmetries, $A_{NC}^{p(n)}$, or the ratios,
$R_{NC}^{{{\bar \nu }/ \nu}, p (n) }$, could be good observables
for the strangeness study
\begin{equation} A_{NC}^{p(n)} =
{ {(\sigma_{NC}^{\nu} - \sigma_{NC}^{{\bar \nu}} )}^{p(n)} \over
{(\sigma_{NC}^{\nu} + \sigma_{NC}^{{\bar \nu}} )}^{p(n)}}~ = {{ 1
- R_{NC}^{{{\bar \nu }/ \nu}, p (n)}} \over { 1 + R_{NC}^{{{\bar
\nu }/ \nu}, p (n) } }} ~,
\end{equation}
where $\sigma_{NC}^{\nu ({\bar \nu}), p(n)}$ means differential
cross sections, Eq.(11), by incident $\nu$ and ${\bar \nu}$ on
proton(neutron).

Since the $y$ variable in Eq.(12) is given as $Q^2 / 2 E_{\nu} M$
in the nucleon rest frame, $y$ is always positive, but less than
$1$ for the energy region, $E_{\nu}< 1 $ GeV and $Q^2 < 1 $
GeV$^2$, considered here. It means that the asymmetry $A_{NC}$
could be very sensitive on the $g_A^s$ value because $A_{NC}$ is
approximated as ${2 y ~ G_M^V G_A }~/~{ ( 1 - y) ( {G_E^V}^2 +
G_A^2) }$ if $O(y^2)$ and ${  E_{\nu} \over { 2 M}} O(y)$ terms
are neglected. Moreover, the Eq.(12) has a positive sign
irrespective of the proton and the neutron. Consequently, $\nu$
cross section is always larger than that of ${\bar \nu}$ on the
nucleon level.

\begin{figure}
\vskip-3.5cm
\includegraphics[width=1.0\linewidth]{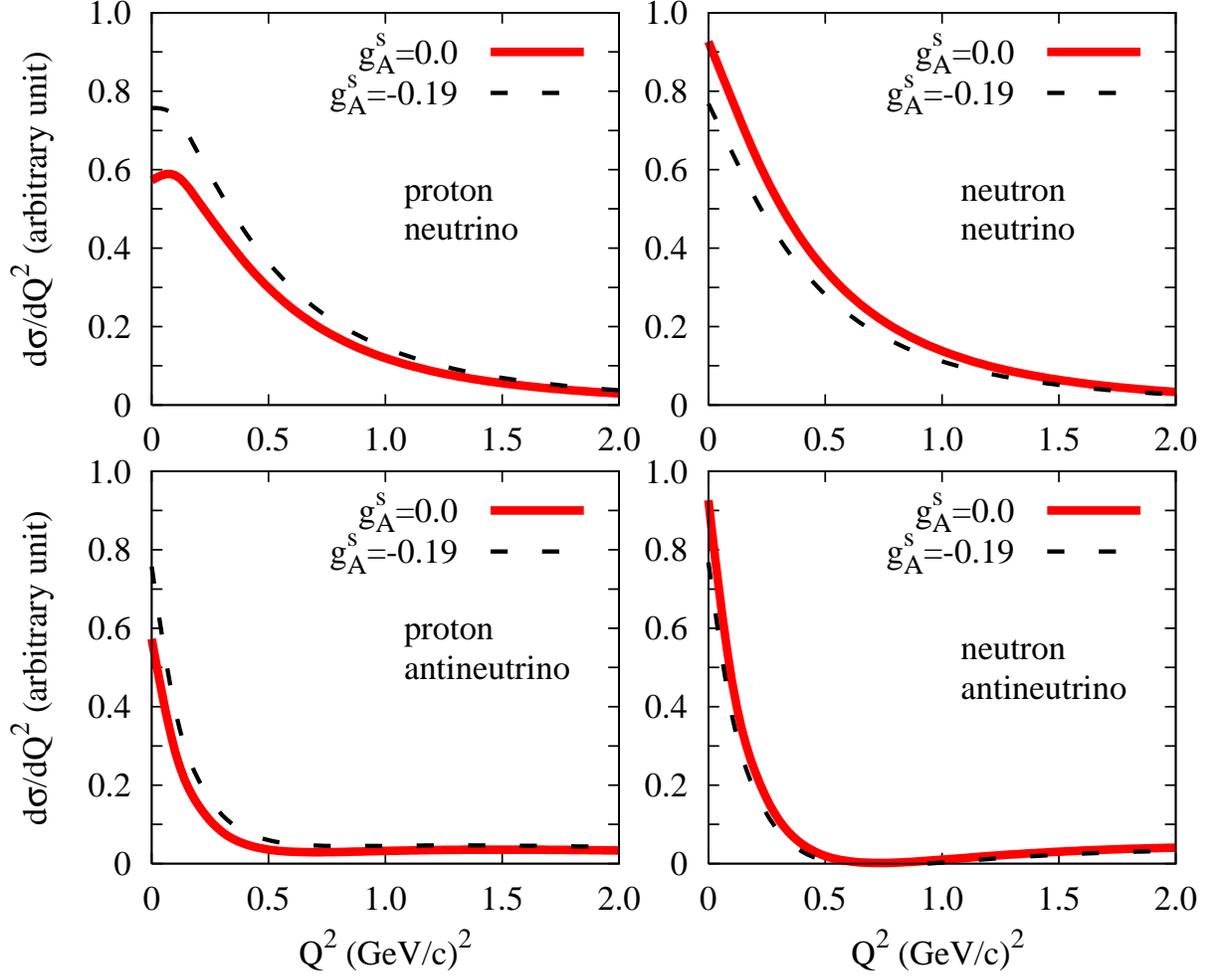}
\caption{Differential cross sections, Eq.(11), by the NC
scattering on proton (left) and neutron (right) in an arbitrary
unit as a function of $Q^2$ for the incident energy $E_{\nu ({\bar
\nu})}=500$ MeV. They are calculated for $g_A^s = -0.19$ and 0.0
cases, respectively. Lower parts are for the ${\bar \nu}$.}
\label{cross}
\end{figure}

\begin{figure}
\vskip-3.5cm
\includegraphics[width=1.0\linewidth]{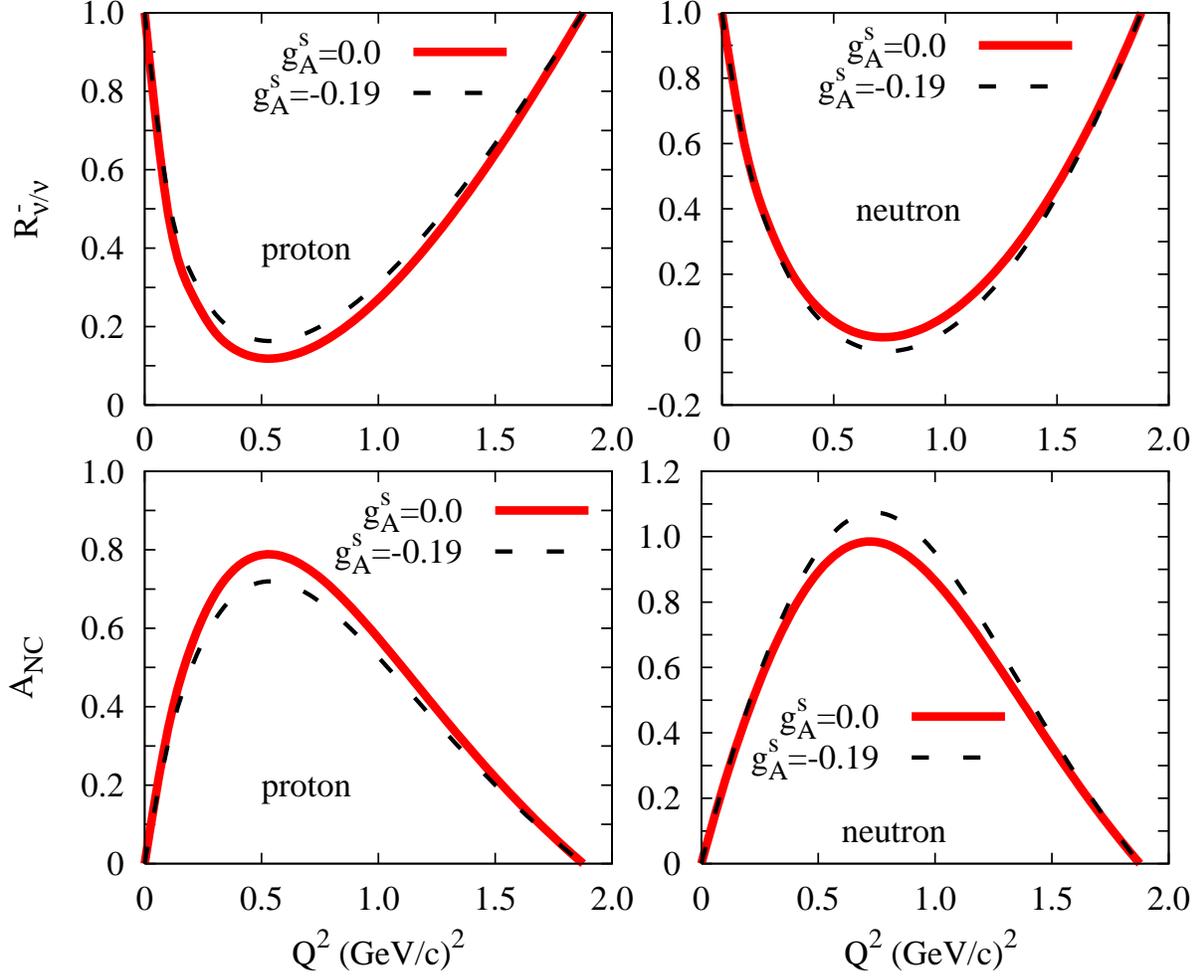}
\caption{$R_{ {\bar \nu } / {\nu} }$ (upper) and asymmetries
(lower), Eq.(14), by the NC scattering on proton (left) and
neutron (right) as a function of $Q^2$ for the incident energy
$E_{\nu}=500$ MeV. They are calculated for $g_A^s = -0.19$ and 0.0
cases, respectively.} \label{asy-rat}
\end{figure}

Detailed results on the nucleon level are shown in Figs.
\ref{cross} and \ref{asy-rat}, where the results for $g_A^s =
-0.19$ and $0.0$ are presented on the proton and the neutron. In
Fig. \ref{cross}, the cross sections by the incident $\nu ({\bar
\nu})$ on the proton are usually enhanced in the whole $Q^2$
region by the $g_A^s$, while they are reduced on the neutron. But,
as shown in Fig.\ref{asy-rat}, the asymmetry on the proton is
maximally decreased in the $Q^2 \sim 0.6 $ GeV$^2$ region about 15
\%, while on the neutron it is maximally increased in that region.
The ratio, $R_{{\bar \nu}/ \nu}$, on the contrary, shows reversed
behaviors. Therefore, the $g_A^s$ effects can be detected in the
asymmetry $A_{NC}$ around the $Q^2 \sim 0.6 $ GeV$^2$ region, more
clearly than the cross sections.

Since the neutrino energy was not known exactly at the BNL
experiments, one usually defines the flux averaged cross section
\begin{equation}
<\sigma> =  {<{ {d \sigma  } \over {d Q^2  } }>}_{\nu ({\bar
\nu})}^{NC} = {  {\int d E_{\nu ({\bar \nu})}  ({ {d \sigma  } /
{d Q^2  }  })_{\nu ({\bar \nu})}^{NC} \Phi_{\nu ({\bar \nu})}
(E_{\nu ({\bar \nu})} )} \over {{\int d E_{\nu ({\bar \nu})}
\Phi_{\nu ({\bar \nu})} (E_{\nu ({\bar \nu})} )}  }} ~,
\end{equation}
where $\Phi_{\nu ({\bar \nu})} (E_{\nu ({\bar \nu})}) $ is
neutrino and antineutrino energy spectra. The experimental result
at BNL, $R_{{\bar \nu} / {\nu}}^{BNL} = {< {\sigma ( {\bar \nu} p
\rightarrow {\bar \nu} p ) }
> / { < \sigma (\nu p \rightarrow \nu  p )> }}$, turned out to be
about 0.32 \cite{brook,Albe02}. Therefore, the flux averaged
asymmetry $< A_{NC} > = { (<\sigma_{NC}^{\nu}> -
<\sigma_{NC}^{{\bar \nu}}> ) / <(\sigma_{NC}^{\nu}> +
<\sigma_{NC}^{{\bar \nu}} > )}$ is 0.5 on the nucleon level. This
value is approximately consistent with our $A_{NC}$ values in
Fig.\ref{asy-rat}, if they are averaged by $Q^2$.

\begin{figure}
\vskip-3.5cm
\includegraphics[width=1.0\linewidth]{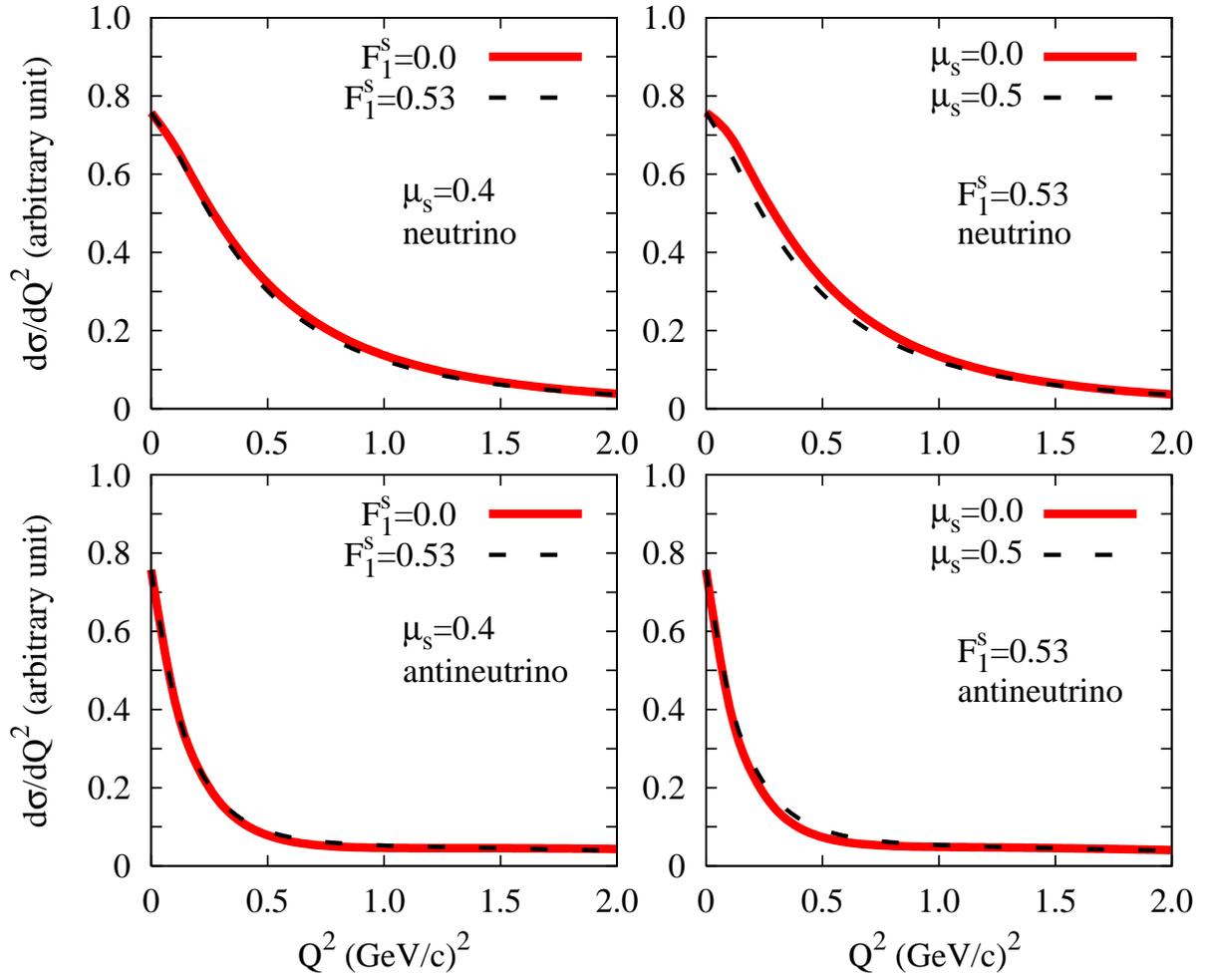}
\caption{$F_1^s$ (right) and $\mu_s$ (left) dependence for the
incident $\nu ({\bar \nu})$ on proton target} \label{f1s-mus}
\end{figure}

Dependence on the strangeness parameters on the vector form
factors, $\mu_s$ and $F_1^s$, is presented in Fig.\ref{f1s-mus}.
Results in left figure are obtained by fixing $\mu_s$ to 0.4, but
varying $F_1^s$ to the constrained interval, $ 0.0 \sim 0.53$.
Those of right panel are the case of $F_1^s = 0.53$ and $\mu_s =
0.0 \sim 0.5$. One can see that only a few percent difference is
found for the cross sections. It means that the most sensitive
effect by the strangeness stems from the axial strangeness form
factor, $g_A^s$, as shown in Figs.\ref{cross} and \ref{asy-rat}.

\begin{figure}
\vskip-3.5cm
\includegraphics[width=1.0\linewidth]{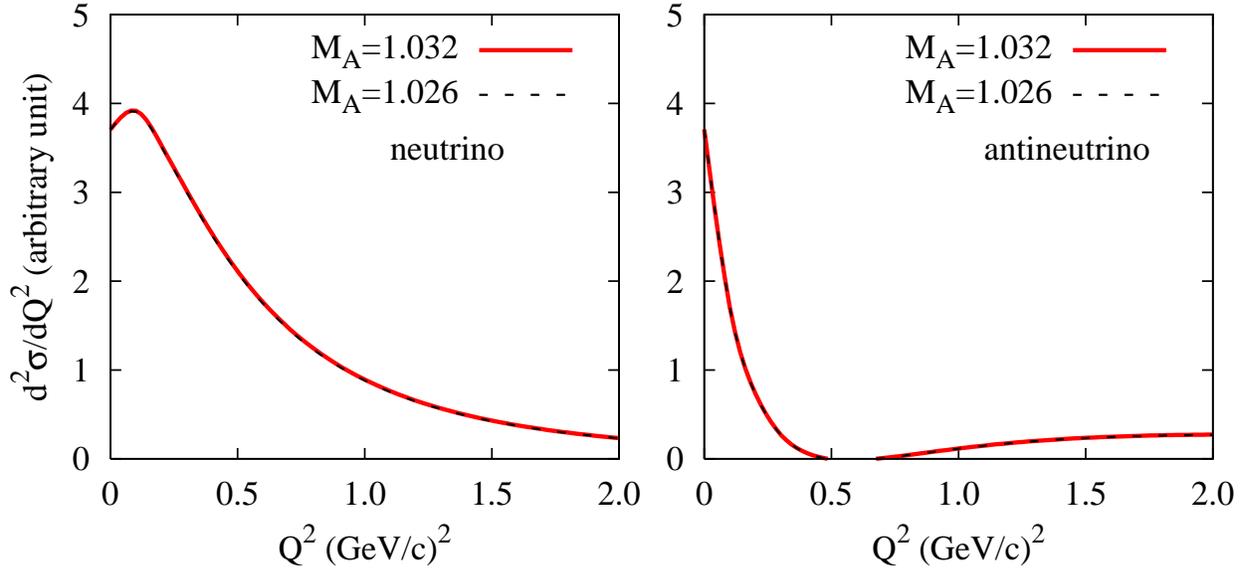}
\caption{Differential cross sections, Eq.(16), for the CC
scattering on the nucleon in an arbitrary unit as a function of
$Q^2$ for the incident energy $E_{\nu ({\bar \nu} )}=500$ MeV.
They are calculated for $M_A$ = 1.032 and 1.026 cases,
respectively. Left panel is w.r.t. $\sigma ( \nu n \rightarrow
\mu^{-} p )$ and right is $\sigma ( {\bar \nu} p \rightarrow
\mu^{+} n )$. Effects of axial mass difference turned out to be
nearly indiscernible.} \label{cross-mass}
\end{figure}

The cross section for CC scattering is given with the following
replacement to NC cross section
\begin{equation}
{( {  {d \sigma  } \over {d Q^2  }} )}_{\nu ({\bar \nu}   )
}^{CC}= {( {  {d \sigma  } \over {d Q^2  }} )}_{\nu ({\bar \nu} )
}^{NC} ( G_E^V \rightarrow G_E^{CC}, G_M^V \rightarrow G_M^{CC},
G_A \rightarrow G_A^{CC})~,
\end{equation}
where
\begin{equation}
G_{E}^{CC} = G_E^p (Q^2) - G_E^n (Q^2)~,~G_{M}^{CC} = G_M^p (Q^2)
- G_M^n (Q^2)~.
\end{equation}
One can see that form factors in CC scattering are not influenced
by the strangeness in the vector and axial form factors, but
depends only on the axial mass $M_A$, and axial coupling constant,
$g_A$. Therefore CC scattering is usually used to extract the
$M_A$ and $g_A$ values \cite{Budd03}. The left and right panel in
Fig.\ref{cross-mass} corresponds to those of $\nu$ and ${\bar
\nu}$ scattering, respectively. The CC cross sections are about 4
$\sim$ 6 times larger rather than those of the NC, and the cross
section of $\sigma_{CC}^{\nu } = \sigma (\nu n \rightarrow \mu^- p
)$ is larger than that of $\sigma_{CC}^{{\bar \nu} } = \sigma
({\bar \nu} p \rightarrow \mu^+ n )$. Our CC results are
consistent with other calculations \cite{Budd03}. Effect of $M_A$
difference, {\it i.e.} cases of $M_A$ = 1.032 and 1.026 $GeV$, is
also presented in Fig.\ref{cross-mass}, but the effect is nearly
indiscernible (see solid and long-dashed curves).

\begin{figure}
\vskip-3.5cm
\includegraphics[width=0.5\linewidth]{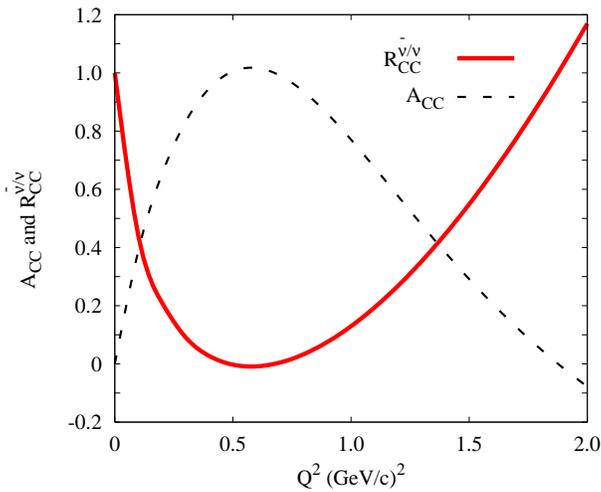}
\caption{Ratio and asymmetry of the CC cross sections in
fig.\ref{cross-mass}} \label{rcc-acc}
\end{figure}

The difference and the sum of $\nu$ and ${\bar \nu}$ scattering
for CC are also given as Eqs.(12) $\sim$ (13) with the above
replacement. If we denote $\sigma_{CC}^{\nu ({\bar \nu})}$ as
differential cross sections by $\sigma (\nu n \rightarrow \mu^- p
)$ and $\sigma ({\bar \nu} p \rightarrow \mu^+ n )$, respectively,
the asymmetry in CC is given as
\begin{equation} A_{CC} = { {(\sigma_{CC}^{\nu} -
\sigma_{CC}^{{\bar \nu}} )} \over {(\sigma_{CC}^{\nu} +
\sigma_{CC}^{{\bar \nu}}  )}}= {{ 1 - R_{CC}^{{\bar \nu }/ \nu}}
\over { 1 + R_{CC}^{{\bar \nu }/ \nu}  }} ~.
\end{equation}
Our results for CC scattering are given in Fig.\ref{rcc-acc},
where dashed and solid curves are the asymmetry and the ratio for
CC, respectively.

\begin{figure}
\includegraphics[width=1.0\linewidth]{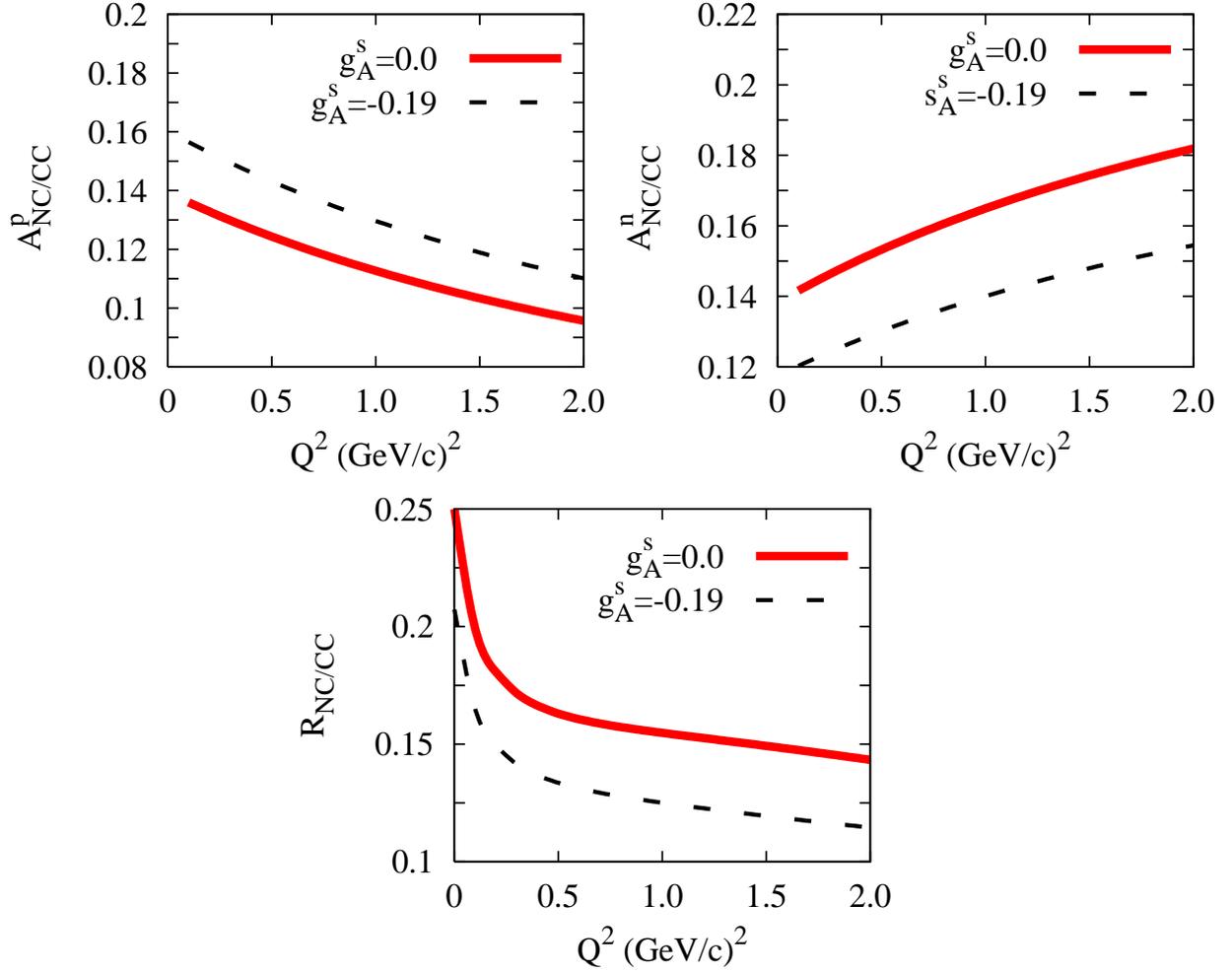}
\caption{Ratio on the neutron, $R_{NC/CC}$ (Eq.(19)), and
asymmetries, $A_{NC/CC}^{p}$ and $A_{NC/CC}^{n}$ (Eq.(20)),
between for NC and CC. They are calculated for $g_A^s = -0.19$ and
0.0 cases, respectively.} \label{rnc-anc}
\end{figure}

Finally we discuss the ratios and the asymmetries between CC and
NC cases. The ratios of NC and CC scattering are given as
\begin{equation}
R_{NC/CC} = {  {\sigma_{NC}^{\nu n} } \over { \sigma_{CC}^{\nu} }}
= {  {\sigma (\nu n \rightarrow \nu n ) } \over {\sigma (\nu n
\rightarrow \mu^- p ) }},~{\bar R}_{NC/CC} = { {\sigma_{NC}^{{\bar
\nu} p} } \over { \sigma_{CC}^{\bar \nu} }} ={ {\sigma ({\bar \nu}
p \rightarrow {\bar \nu} p )  } \over {\sigma ({\bar \nu} p
\rightarrow \mu^+ n ) }}~.
\end{equation}
These ratios have been suggested for probing the strangeness on
the nucleon or nuclei because the CC scattering is independent of
the strangeness and any possible nuclear structure effects are
expected to be cancelled out. Results for $R_{NC/CC}$ are shown in
Fig.\ref{rnc-anc}. One can see large strangeness effects due to
the $g_A^s$ in the ratios (see solid and dashed curves). Here we
$F_1^s = 0.53$ and $\mu_s = -0.4$ are used.

It would be interesting which is the larger of the two strangeness
effect, vector and axial strangeness parts. To distinguish each
strangeness contribution, we consider asymmetries between NC and
CC, because they are expressed only in terms of the ratios between
strangeness and non-strangeness as follows
\begin{eqnarray}
A_{NC/CC}^p & = & { {\sigma_{NC}^{\nu p} - \sigma_{NC}^{{\bar \nu}
p} } \over {(\sigma_{CC}^{\nu n} - \sigma_{CC}^{{\bar \nu} p}  )}}
= { {G_M^{V,p} G_A   }\over {G_M^{CC} G_A^{CC}  }} = 0.12 - 0.12 {
{g_A^s } \over { g_A }}
- 0.13 { {G_M^s } \over {G_M^3  }}~,\\
\nonumber A_{NC/CC}^n & = & { {\sigma_{NC}^{\nu n} -
\sigma_{NC}^{{\bar \nu} n} } \over {(\sigma_{CC}^{\nu n} -
\sigma_{CC}^{{\bar \nu} p} )}} = { {G_M^{V,n} G_A   }\over
{G_M^{CC} G_A^{CC}  }} = 0.16 + 0.16 { {g_A^s } \over { g_A }} +
0.13 { {G_M^s } \over {G_M^3  }}~,
\end{eqnarray}
where $G_M^s ( Q^2 = 0) = \mu_s$, $G_M^3 = { (G_M^p - G_M^n}) /
2$, $\mu_p = 2.79$, and $\mu_n = -1.91$ are used. Since $ \vert
G_M^s/G_M^3 \vert $ and $ \vert g_A^s/g_A \vert$ are approximately
0.2, strangeness effects from vector and axial parts are
comparable, in principle.

Remarkable point is that one can decide the $\mu_s$ sign. If
$g_A^s$ and $G_M^s$ have different $\pm$ sign, $A_{NC/CC}^{p (n)}
$ values become constants 0.12, 0.16, respectively, because the
last two terms in Eq.(20) are nearly cancelled. Any deviations
from these constants by the same sign would imply the $Q^2$
dependence of form factors as shown in Fig.\ref{rnc-anc}, in which
we used $g_A^s = -0.19$ and $\mu_s = -0.4$.

\begin{figure}
\includegraphics[width=0.7\linewidth]{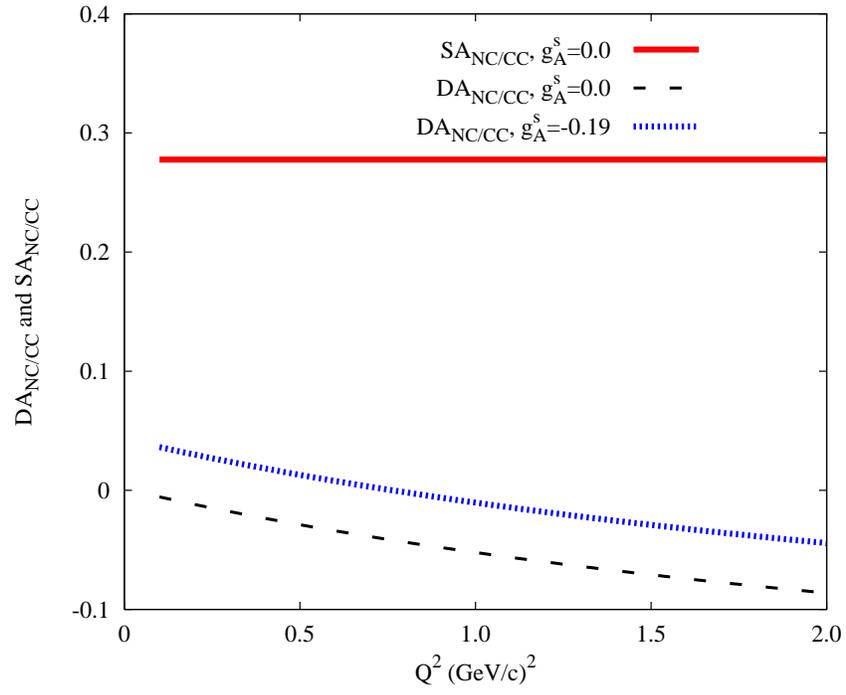}
\caption{Difference and sum, Eq.(21), of asymmetries in
Fig.\ref{rnc-anc}} \label{sa-da}
\end{figure}

Finally we consider a difference and a sum of asymmetries
\begin{eqnarray}
DA_{NC/CC}^{p,n} & = & A_{NC/CC}^p - A_{NC/CC}^n ={( {G_M^{V, p} -
G_M^{V,n} )G_A } \over { {G_M^{CC}  G_A^{CC} }} }\simeq - 0.04 -
0.28 ({
{g_A^s } \over { g_A }} + { {G_M^s } \over {G_M^3  }})~,\\
\nonumber SA_{NC/CC}^{p,n} & = & A_{NC/CC}^p + A_{NC/CC}^n = {(
{G_M^{V, p} + G_M^{V,n} )G_A } \over { {G_M^{CC} G_A^{CC} }} }=
0.28 + 0.04 { {g_A^s } \over { g_A }} ~ .
\end{eqnarray}
Sum of asymmetry, $SA_{NC/CC}^{p,n}$, is given only in terms of
axial part. But the 2nd term is very small by the factor 0.04, so
that SA is nearly independent of the strangeness, but DA depends
strongly on the axial strangeness as shown in Fig.\ref{sa-da}. If
$g_A^s$ and $\mu_s$ have different signs, DA would be constant,
but it depends on $Q^2$ if they have same signs as in
Fig.\ref{sa-da}.

In the following, we make brief summaries and conclusions. A newly
combined data by parity violating electron scattering and $\nu
({\bar \nu})$ scattering shed a valuable light on relevant
strangeness form factors, $G_{E,M}^s ( Q^2)$ and $G_A^s (Q^2)$.
Using a conventional dipole form, we extracted lower and upper
limits of the parameters related with the strangeness for factors
by adjusting them to the experimental data. We found that $G_M^s
(Q^2 = 0) = \mu_s$ has a positive value contrary to the negative
values exploited for $\nu - A$ scattering calculations.

Effects of the strangeness form factors are investigated for cross
sections and their ratios by varying the parameters within the
limits. We found ambiguities by $F_1^s , \mu_s$ due to the
vectorial and $g_A^s$ by the axial form factors are within
maximally 3 \%, 5 \%, and 15 \%, respectively. Change of axial
mass in these calculations affect results within only a 1 \%.

In specific, effects of the main parameters, $g_A^s$, are
detailed. It shows that cross sections, ratios and asymmetries for
protons are increased, decreased, and increased, respectively, by
the $g_A^s$, while those of neutrons show reversed behaviors of
each quantity.

Finally, in order to search for more efficient observable for the
strangeness, relevant asymmetries between $\nu$ and ${\bar \nu}$
scattering are studied. As expected, there appeared larger
strangeness effects in the ratios and asymmetries rather than
cross sections. In specific, asymmetry between NC and CC
scattering could be meaningful methods to look for the $Q^2$
dependence of the strangeness form factors. It is also remarkable
that sum of asymmetry is nearly independent of the strangeness.
Therefore it could be a measure of nuclear effects independent of
the strangeness. Nuclear application based on this work are under
progress. One of our preliminary results show that $g_A^s$ effects
are strongly cancelled in nuclei by the enhancement of proton
knockout processes and the decrease of neutrons processes.

{\bf Acknowledgments}

This work was supported by the Soongsil University Research Fund.

\newpage
\begin{table}
\caption[Coupling constants and parameters to the strangeness form
factors. ] {\small Relevant parameters to the axial couplings and
the strangeness form factors. } \vskip 0.2cm
\setlength{\tabcolsep}{2.0mm} {\scriptsize
\begin{tabular}{|c|c|c|c|c|c|c|c|}\hline
$\sin^2 \theta_W$  & $g_A$ & $M_A$ & $g_A^s = \Delta s$ &
${F_1^s}^{\dagger}$ & $\mu^s = F_2^s (0)$ & {$\rho^s$}$^{**}$ &
Ref.
 \\ \hline \hline
   0.23143 & 1.26 & 1.026 $\pm$ {0.021} \cite{Bern02}
   &  -0.10 & 0.4 & -0.50 & 2 &  \cite{giusti1} (2006)
                                  \\ \hline
   0.2313  & 1.26 & 1.026 $\pm$ {0.021} \cite{Bern02} & -0.19
   & 0.53 & -0.40 & x & \cite{giusti1} (2004)
                                  \\ \hline
 0.2224  & 1.262 & 1.032 & x
   & x & x & x & \cite{udias}
                                   \\  \hline
 0.2325  & 1.256 $\pm$ 0.003 & 1.012 $\pm {0.032}^{*}$ & -0.21
 $\pm$ 0.10
   & 0.53$\pm$ 0.70 & -0.40 $\pm$ 0.72 & x & \cite{garvey} (Model
   IV)
   \\ \hline
    0.2325  & 1.256 $\pm$ 0.003 & 1.049 $\pm$ {0.023}  & -0.13
 $\pm$ 0.09  & 0.49 $\pm$ 0.70 & -0.39 $\pm$ 0.70 & x & \cite{garvey} {(Model
 III)}
\\  \hline
    x  & x & 1.026 $\pm$ {0.021} \cite{Bern02} & -0.04 $\sim$ -0.09
   & x &  0.08 $\sim$ 0.32& x & Theory \cite{Step06,Silvia,Riska}
\\  \hline
0.232  & 1.256 & 1.032 & -0.21 $\sim$ 0.0
   & 0 $\sim$ 0.53 &  0.0 $\sim$ 0.4 $^{\dagger \dagger}$ & x & Ours
\\  \hline
\end{tabular}
} \label{kthT4}
\end{table}
$^*$ a value cited from a table at ref. \cite{garvey}, but
actually the world average value prior to the BNL data, 1.032
$\pm$ 0.036, seems to be used at the reference.

$^{\dagger}$ $F_1^s=-<r_s^2>/6$ is the rms value of strangeness.

$^{**}$ comes from ref. \cite{giusti1} in which a bit different
form for $F_{i}^s (Q^2)$ with a constant $\rho^s$ is used.

$^{\dagger \dagger}$ has a positive sign (see the data at ref.
\cite{Step06}) compared to previous values.

\newpage

\newpage

\newpage

\newpage

\end{document}